\documentclass{emulateapj}
\usepackage{graphicx}
\usepackage{epsfig}
\shorttitle{Designer Cosmology}
\shortauthors{Bassett, Parkinson \& Nichol}
\begin{document}
\def\beq{\begin{equation}}
\def\eeq{\end{equation}}
\def\bea{\begin{eqnarray}}
\def\eea{\end{eqnarray}}
\def\thet{\theta_{\mu}}
\def\Sig{\Sigma_{\mu}}
\def\eps{\epsilon_i}
\def\s{${\bf s }$}
\def\T{${\bf \Theta}$}
\title{Designer Cosmology}
\author{Bruce A. Bassett$^{1,2}$, David Parkinson$^{2}$ and Robert C. Nichol$^{2}$}
\affil{$^1$ Department of Physics, Kyoto University, Kyoto, Japan}
\affil{$^2$ Institute of Cosmology and Gravitation, 
University of Portsmouth, Portsmouth~PO1~2EG, UK}
\begin{abstract}
The IPSO framework allows optimal design of experiments and surveys. We discuss 
the utility of IPSO with a simplified 10 parameter 
MCMC D-optimisation of a dark energy survey. The resulting optimal number 
of redshift bins is typically two or three, all situated at $z<2$. By 
exploiting optimisation we show how the statistical power of the survey 
is significantly enhanced.  Experiment design is aided by the richness 
of the figure of merit landscape which shows strong degeneracies, 
which means one can impose secondary optimisation 
criteria at little cost. For example, one may choose either to maximally test a single model (e.g. $\Lambda$CDM) 
or to get the best model-independent constraints possible (e.g. on a whole space of dark energy models). 
Such bifurcations point to a future where cosmological experiments become increasingly 
specialised and optimisation increasingly important.
\end{abstract}
\keywords{cosmological parameters --- large-scale structure of universe --- 
surveys}
\section{Introduction}
We have reached an enviable resonance in which improvements in detector 
performance and cost are allowing not only rapid gains in our fundamental 
knowledge of the cosmos but also the opportunity for smaller experiments 
to make critical contributions to that knowledge. This has resulted in a 
surge of interest in next-generation experiment design with over twenty 
major surveys in planning or construction in observational cosmology alone. 
Experimental cosmology has changed in a few short years into a crowded and 
jostling marketplace.  

There are several big prizes currently at stake: {\em Detection of dark 
energy dynamics, B-mode polarisation and cosmological non-Gaussianity}. 
Competition, limited funding, low signal-to-noise and extreme competition 
mean that new surveys will need to be increasingly optimised to get the 
most out of them. The aims of this {\em Letter} are to show how this can 
be achieved in a cross-disciplinary way and to illustrate some of the rich 
aspects of cosmological optimisation.

\section{IPSO}
Integrated Parameter Space Optimisation (IPSO; Bassett 2004, hereafter B04) 
proceeds by first constructing a class of candidate survey/experiment 
geometries, $S$, labeled by survey parameters, $s_i$, such as areal and redshift coverage. 

Second, a target parameter space, \T, is defined, consisting of the 
parameters that we wish to optimally constrain (labeled 
$\theta_{\mu,\nu...}$). There are also typically nuisance parameters 
we need to marginalise over (labeled $\varphi_{a,b...}$). 

Third, a Figure of Merit (FoM) is defined which assigns a single real 
number to each candidate survey. The candidate with the extremal FoM is the 
optimal experiment/survey. The FoM we consider is defined by (B04):
\beq
FoM(s_i) = \int_{\bf \Theta} I(s_i,\vec{\theta}) p(\vec{\theta}) d\vec{\theta}\,.
\label{basicdef}
\eeq
$I(s_i,\vec{\theta})$ is a scalar which depends on the survey geometry (through the $s_i$), 
and position in ${\bf \Theta}$ and 
$p(\vec{\theta})$ is a `window function' that weights the different 
regions of the parameter space. By integrating over the parameter space we do not 
make assumptions about the underlying model, which is particularly 
important when we have very limited knowledge 
of the underlying physics, as is the case with dark energy.

Most choices for $I(s_i,\vec{\theta})$ typically invoke either the 
parameter covariance matrix or ${\bf F}$, the Fisher matrix, defined by:
\beq
F_{AB} = -\left\langle \frac{\partial^2 {\cal \ln L}}{\partial\theta_A\partial\theta_B}\right\rangle = \sum_i \left(\frac{\partial X}{\partial \theta_A}\frac{\partial X}{\partial \theta_B}\right)_i\epsilon_i^{-2}(s)\,.
\label{fisher1}
\eeq
Here we use $A = \{\mu...,a...\}$ to label both fundamental and nuisance 
parameters, ${\cal L}$ is the likelihood, $X = C_{\ell}, d_L, H$ represents 
the quantity being measured with $i$ labeling redshift bin or Fourier mode 
as appropriate (Tegmark {\it et al.} 1998). The $\epsilon_i^2$ are the 
error variances on $X$ and depend explicitly on the survey parameters, $s_i$, 
unlike the derivatives, $\partial X/\partial \theta_A$. In computing integrals such as (\ref{basicdef}) 
this allows for significant cpu gains since the derivatives need only to 
be computed once. 

Via the Cram\'er-Rao bound ${\bf F}^{-1}$ provides the best possible 
covariance matrix and hence a lower bound on the achievable parameter 
variances.  Although there are many choices for $I(s_i,\thet)$ (B04) we focus on only 
one for simplicity: D-optimality, defined by
\beq
I(s,\thet)  =  \log \mbox{det}({\bf F} + {\bf P}) ~~~~~{\bf D-optimality} \label{dop}
\eeq
where `det' denotes matrix determinant and 
${\bf P}$ is the prior precision matrix, {\em viz.} the Fisher matrix of all the relevant prior data.

\begin{figure}
\plotone{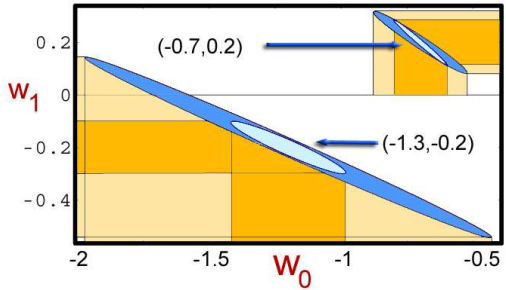}
\caption[survey]{{\bf Typical D-optimality improvements on error ellipses} 
at the two base points  $(w_0,w_1) = (-1.3,-0.2)$ and $(-0.7,0.2)$. The unoptimised 
survey (outer dark ellipses) has errors between $50\%$ and three times 
larger in both $w_0$ and $w_1$ than the D-optimal survey (inner light  
ellipse) which was optimised around $\Lambda$CDM. Note the 
particularly significant gains in the phantom region $w_0 < -1$.}
\label{phantom}
\end{figure}

Eq. (\ref{dop}), is the gain in Shannon information or entropy over the prior. 
Maximising (\ref{dop}) provides the best possible gain in 
constraints on the parameters $\thet$ over what was available 
from just the prior data, ${\bf P}$. It is known as 
D-optimality in the design literature. If ${\bf P}=0$
maximising (\ref{dop}) is equivalent to minimising the volume of the error ellipses, an
alternative FoM (Huterer and Turner 2001, Frieman {\em et al.} 2003, B04). 
Via the General Equivalence Theorem, D-Optimal solutions are also optimal
under other FoM. For these reasons it seems appropriate for cosmological 
applications, although as we will see, secondary optimisation criteria can be imposed at 
almost no cost to the primary FoM. 

Nuisance parameters, such as $\Omega_k, \Omega_m$ etc..., whose values we do not know precisely but which 
we do not want to optimise with respect to, can be easily dealt with by 
inverting the full Fisher matrix $F_{AB}$, extracting 
the relevant submatrix corresponding to the $\thet$, re-inverting (e.g. Seo \& Eisenstein 2003, B04) and 
then applying Eq. (\ref{dop}). Further, any reasonable FoM can also be generalised to allow inclusion of competing 
surveys by simply replacing ${\bf F} \rightarrow {\bf F} + {\cal F}$ where 
${\cal F}$ is the sum of the Fisher matrices expected for the competing 
surveys. In this way IPSO will find the optimal niche with respect to the 
other surveys (B04).

\section{Optimizing CMB and Weak Lensing Surveys} \label{wl}

When is optimisation worth doing? To illustrate this let us contrast  
weak lensing (wl) convergence and CMB surveys on the celestial sphere. In both of these cases the Fisher 
matrix is a sum over $\ell$ (Hu and Tegmark 1999, Knox and Song 2002, Kesden {\em et al.} 2002):
\beq 
F_{\mu\nu} =  \sum_{l>f_{sky}^{-1/2}}\frac{2\ell+1}{2} (N^X_{\ell})^{-2} \frac{\partial C_{\ell}^X}{\partial \theta_{\mu}}\frac{\partial C_{\ell}^X}{\partial \theta_{\nu}}
\eeq
where $X=CMB,wl$, $N^X_{\ell}$ is the total noise for the survey and $f_{sky}$ is the fraction of the sky observed. 
For the CMB we consider only one spectrum (e.g. the B-mode power spectrum). 
In both cases we assume that the surveys are constrained to last a given 
length of time, $T$, and ask `what is the optimal sky coverage, $f_{sky}$, 
given this constraint?' For CMB experiments we have (Knox 1995):
\beq
N^{CMB}_{\ell} \propto  f_{sky}^{-1/2}(C^{CMB}_{\ell} + \frac{a f_{sky}}{T} e^{\ell^2 \sigma_b}) 
\label{N}
\eeq 
where $T = t_{pix}N_{pix}$ is the length of the survey, 
$a$ is a proportionality constant and the $N_{pix}$ pixels are each observed for 
time $t_{pix}$ using a Gaussian beam with $\mbox{FWHM} \propto \sigma_b$.  
The first (second) term in (\ref{N}) is the noise from sample variance (instrument noise). 

CMB experiments will benefit from optimisation since the competition between the terms in eq. (\ref{N}) creates a local minimum in the noise (Jaffe {\em et al} 1999). To apply IPSO to the CMB one must first choose \T. For example, for optimal detection of deviations from the inflationary consistency conditions the key variable is $\theta \equiv n_t + r/4.8$ where $n_t$ is the tensor spectral index and $r$ is the ratio of tensor to scalar  quadrupole in the CMB. Single field inflation predicts this should vanish. Hence a high-$\sigma$ detection of $\theta \neq 0$ would put severe pressure on simple inflationary models. In contrast, an experiment designed to detect B-mode polarisation alone would optimise to detect $r$ only and would lead to a different optimal area. 

In contrast, for weak lensing (Kaiser 1992)
\beq
N^{wl}_{\ell} \simeq f_{sky}^{-1/2}C^{wl}_{\ell} + \frac{\sigma_g^2}{2\sqrt{T}\overline{n}}
\label{nwl}
\eeq
where $\sigma_g^2 \sim 0.35$ is the approximately constant intrinsic ellipticity error and the surface density of detected galaxies scales roughly as $\overline{n} \sqrt{t}$ where $t$ is the integration time per field of view. The noise terms $N^X_{\ell}$ differ crucially when it comes to optimisation of the areal coverage, $f_{sky}$.  Unlike the CMB noise, $N^{wl}_{\ell}$ has no local minimum; the weak lensing Fisher matrix is a monotonic function of $f_{sky}$. Optimising any of the FoM simply proceeds by using the largest feasible area to minimise the sample variance. 

If, in addition, the intrinsic ellipticity noise dominates the noise (as it does for the proposed SNAP WAS) then the FoM becomes essentially independent of $f_{sky}$ and the gain of going to the largest area is minimal, as found by (Rhodes {\em et al.}, 2004). 

\section{Optimal measurements of the Hubble constant}\label{hub}

To illustrate some of the issues one faces in applying IPSO to realistic surveys, consider 
the optimisation of a redshift survey designed to measure 
the Hubble constant through observation of the radial baryon oscillations 
(Seo \& Eisenstein 2003, Blake \& Glazebrook 2003, Linder 2003, Amendola {\em et al.} 2004, Yamamoto {\em et al.} 2005).  For clarity we assume no nuisance parameters, a flat FLRW model with $\Omega_m = 0.3$, $H_0$ known exactly 
and we ignore the constraints from $d_A$ which a full optimisation would include. 

We consider a model of dark energy based on Taylor expansion in powers of $(1-a) = \frac{z}{1 + z}$ 
(Chevallier M., Polarski 2001, Linder 2003, Bassett, Corasaniti \& Kunz 2004), with $w = p_{DE}/\rho_{DE}$ 
\bea
w(z) &=& w_0 + w_1\frac{z}{1+z} + w_2 \frac{z^2}{(1+z)^2}\nonumber\\ 
\rho_{DE} &\propto& (1+z)^{3(1 + w_0 + w_1 + w_2)} \times \nonumber\\ 
& &  \exp\left(\frac{-3z(2w_1(1 + z) + w_2(2 + 3z))}{2(1+z)^2}\right)
  \label{rhode}
\eea
where $\rho_{DE}(z)$ is the dark energy energy density. 

\begin{figure}
\plotone{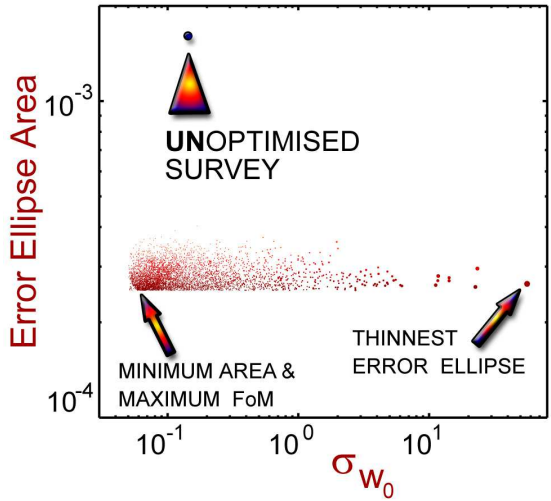}
\caption[survey]{
{\bf Error on $w_0$ ($\sigma_{w_0}$) versus area of the error ellipse}. Note the 
wide range of $\sigma_{w_0}$ at almost constant minimum area. 
$\sigma_{w_0}$ effectively measures the ellipticity of the error ellipse. 
At nearly constant FoM one can optimise to obtain circular or very thin ellipses, depending on 
one's aims. The size of the points is proportional to the error in $w_1$.
\label{w0vol}}
\end{figure}

Comparing with eq. (\ref{fisher1}), $X=E\equiv H/H_0$. Rather than the optimal area, $f_{sky}$, we want the optimal number of redshift bins, $N$, what redshifts they should be centered on, $z_i$, and how long we should observe in each bin, $t_i$. Again we assume fixed total survey time ($T$) so we need to optimise given the constraint $\sum_i^N t_i = T$. 

We assume that the error bars scale as $\epsilon_i^{-2} = e_i t^{\gamma}$ where $e_i \propto (1+z_i)^{-\beta}$ gives the efficiency with which galaxies are detected and $\gamma, \beta$ parameterise our ignorance. These could be treated as nuisance parameters to be marginalised over but we find that our main results are insensitive to both over the range $\beta = 1-2, \gamma=1-2$ we consider. Note that $\gamma=1$ implies that the FoM is maximised on the boundary of the allowed redshift region (just as was the case with the weak lensing survey earlier). We focus on the case $\gamma=2$ here for illustrative purposes. The real constraint will be significantly more complex and we leave this issue to future work. 

We restrict the bin redshifts to be between $0.5 < z_i <4.5$ which is a feasible range for future baryon oscillation surveys such as KAOS and set $w_2 = 0$ for clarity. We performed an MCMC (Christensen and Meyer, 2000) optimisation with the D-Optimal FoM (\ref{dop}) with ten free survey parameters: $\{t_i, z_i\}$ giving the integration time and redshifts of the five bins.  The effective number of bins varies dynamically because the MCMC chain could (and typically did) assign negligible amounts of observing time to some of the bins.

We ran multiple chains (up to 5000) with random starting configurations for the survey and used a standard Hastings-Metropolis algorithm for jump acceptance. Instead of directly performing the integral (\ref{basicdef}) we used the definition $FoM(s) = \frac{1}{N} \sum_{i=1}^N I(s,\theta_a)$ where the $\theta_a$are drawn randomly from a probability distribution based on $p(\theta_i)$ in (\ref{basicdef}). We chose $p$ to be a bi-variate Gaussian centered on the $\Lambda$CDM point $w_0=-1, w_1 = 0$ so our optimisation was chosen to detect slowly varying dark energy dynamics close to a cosmological constant.

\begin{figure}
\plotone{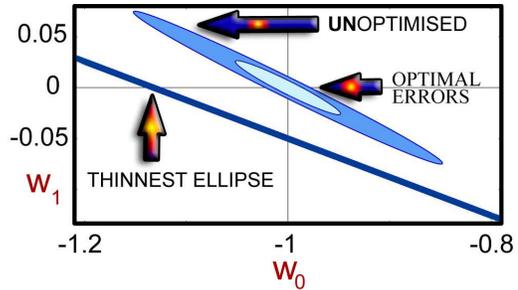}
\caption[survey]{
{\bf Choices in optimisation}. At constant FoM one can 
optimise to achieve thin error ellipses with tiny transverse errors 
(diagonal dark blue ``strip") or to achieve the best joint constraints on 
all parameters (light blue ellipse). All ellipses are computed at $w_0 = -1, w_1 = 0$ 
and reduce the total unoptimised ellipse area (dark blue ellipse) by about $600\%$. 
\label{choice}
\label{lcdm}}
\end{figure}

\begin{figure}
\plotone{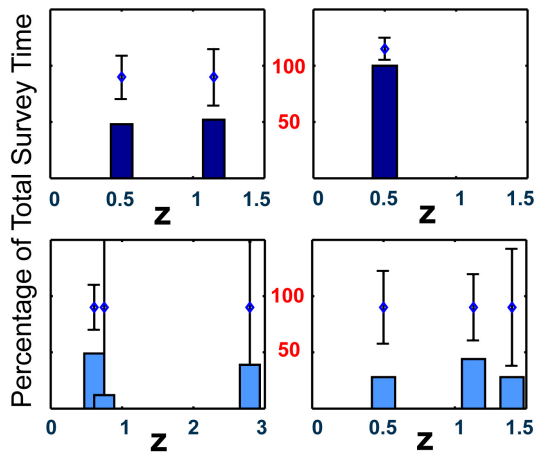}
\caption[survey]{
{\bf Redshifts and integration times for the optimal surveys 
shown in Fig. (\ref{w0vol}) and the resulting error bars on $H(z)$. }
{\bf Top Left} - D-optimal maximum FoM (which 
also has the minimum error ellipse area). It splits the total survey equally 
between $z=0.5$ and $z = 1.15$. {\bf Top Right} - the thinnest possible 
ellipse with all measurements at a single redshift, $z = 0.5$. The 
{\bf bottom two panels} show the geometries for the 2nd (left) and 
2000th (right) largest FoM (only 0.005\% and 6\% smaller than the maximum 
respectively). This shows the diversity of geometries with nearly 
degenerate FoM. 
\label{survey}}
\end{figure}

The Fisher matrix derivatives based on eq. (\ref{rhode}) are simple to compute; e.g. 
\bea
\frac{\partial \rho_{DE}}{\partial w_{0}} &=& 3 \rho_{DE}~ \ln(1+z)\nonumber\\
\frac{\partial \rho_{DE}}{\partial w_{1}} &=& 3 \rho_{DE} ~(\ln(1+z) - \frac{z}{1+z} )\,.
\eea  
Our unoptimised fiducial survey had five redshift bins located at $z_i = 0.6,0.8,1,1.2,3$, as in (Seo \& Eisenstein 2003 and Amendola {\it et al.} 2004), with equal integration time ($T/5$) assigned to 
each bin. 

Fig. (\ref{phantom}) shows typical gains over the unoptimised survey for a near optimal survey chosen randomly from the 5000 MCMC chains while Fig. (\ref{w0vol}) shows the area of the error ellipse versus the corresponding error on $w_0$ for each of the 5000 locally optimal solutions. It is very clear that at almost identical area (and FoM) there is a very wide range of error ellipse ellipticity (controlled by $\sigma_{w_0}$).  In other words, there are many local maxima which come very close to matching the global maximum.

The implications of this FoM `degeneracy' for ruling out dark energy models are clarified in Fig. (\ref{lcdm}) where we show the thinnest error ellipse (diagonal ``strip"), the unoptimised error ellipse and the error ellipse with the maximal FoM, all computed at the $\Lambda$CDM point (the thinnest ellipse is shifted down for clarity). 

This degeneracy offers the chance for secondary optimisation (the primary one in this case being based on the D-optimal FoM). For example, one could choose geometries that deliver the best constraints on a particular linear combination of the parameters $\thet$ adapted to the degeneracy structure of the observations while sacrificing the orthogonal direction(s). This amounts to minimising the smallest eigenvalue of the covariance matrix which may be preferable for testing dark energy dynamics in the short term. 

\begin{figure}
\plotone{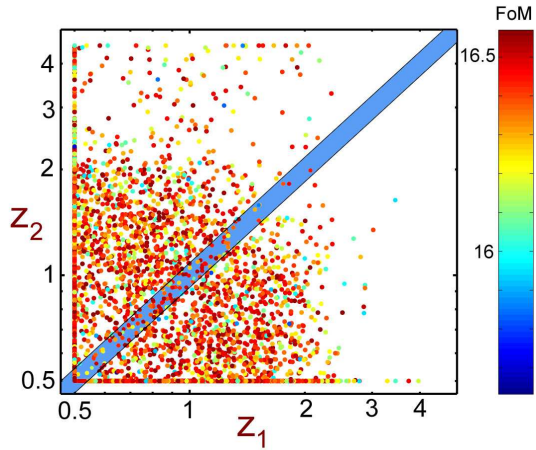}
\caption[survey]{Phase plot of the redshifts, $z_1,z_2$, of the two dominant bins with the 
most integration time for 5000 separate MCMC optimisation chains. The dominant bins 
are typically found at $z_{1,2} < 2$ although surveys with $z_2 > 2$ can be competitive. 
Note the clustering around the minimum redshift we considered, $z = 0.5$. 
The diagonal band results in very thin error ellipse 
(large $\sigma_{w_0}$ in figure (\ref{w0vol}).
\label{domz}}
\end{figure}

The redshifts, $z_i$, and integration times, $t_i$, for each bin of some optimal and near optimal surveys are shown in Fig. (\ref{survey}) along with corresponding error bars on the Hubble rate, $H(z)$. Typically the locally optimal geometries in our 5000 chains had only two ($51\%$ of all chains) or three redshift bins ($39\%$ of all chains) with more than $5\%$ of the total survey time. Optimal geometries with either one or five bins were extremely rare, forming less than $1\%$ of all the locally optimal geometries (although single bin geometries deliver the thinnest error ellipses). The preferance for only a few bins  arises because the dark energy models we consider vary rather slowly with redshift, hence it is statistically preferable to constrain $w$ rather than $dw/dz$. This conclusion may change somewhat if one allows very rapid evolution in $w(z)$ which actually provides a very good fit to current SNIa data (Bassett {\em et al}., 2004). We found that the redshifts of the two bins with the most integration time were typically located at $z<2$ (as shown in Fig. (\ref{domz})). 

\section{Conclusions}
We have considered optimisation both of 2-d surveys such as CMB and weak lensing experiments and 3-d redshift surveys. In a simplified optimisation of a baryon oscillation survey we have shown how IPSO allows significant gains in the statistical power of a survey can be achieved through optimisation, in this case a reduction by a factor of 6 in the error ellipse area over the unoptimised survey. We found that there are many diverse surveys with nearly degenerate figures of merit (FoM), as shown in Figs. (\ref{survey}) \& (\ref{w0vol}). This is good news since it allows survey designers to pick a near optimal survey structure that is most compatible with real-world intangibles that cannot easily be included explicitly in the optimisation. 

The Monte Carlo Markov Chain (MCMC) search was repeated thousands of times with randomly chosen initial survey configurations. Most of the resulting locally optimal surveys divided $> 90\%$ of the survey time between only two or three redshift bins. A single bin leads to the thinnest possible error ellipse and may be appropriate for some experiments, particularly if the resulting ellipse is orthogonal to those coming from other observations. 
Alternatively, at almost the same FoM, one can choose a survey configuration that gives the best joint constraints on all the parameters simultaneously.  At least for measurements of the Hubble constant alone, we found that typically the two dominant redshift bins should be located at low redshift, $z<2$, as shown in Figures (\ref{survey}) and (\ref{domz}). This is good news for upcoming baryon oscillation surveys such as KAOS which will be able to probe the optical region $z<1.3$ with high precision from earth. 

\acknowledgments

We thank Chris Blake, Eric Linder and Takahiro Tanaka for useful comments on the draft.

\end{document}